# Routing Autonomous Emergency Vehicles in Smart Cities Using Real Time Systems Analogy: A Conceptual Model

Subash Humagain✉ and Roopak Sinha
*IT and Software Engineering*
*Auckland University of Technology, Auckland, New Zealand*
*subash.humagain@aut.ac.nz, roopak.sinha@aut.ac.nz*

**Abstract**

*Emergency service vehicles like ambulance, fire, police etc. should respond to emergencies on time. Existing barriers like increased congestion, multiple signalized intersections, queued vehicles, traffic phase timing etc. can prevent emergency vehicles (EVs) achieving desired response times. Existing solutions to route EVs have not been successful because they do not use dynamic traffic parameters. Real time information on increased congestion, halts on road, pedestrian flow, queued vehicles, real and adaptive speed, can be used to properly actuate pre-emption and minimise the impact that EV movement can have on other traffic. Smart cities provide the necessary infrastructure to enable two critical factors in EV routing: real-time traffic data and connectivity. In addition, using autonomous vehicles (AVs) in place of normal emergency service vehicles can have further advantages in terms of safety and adaptability in smart city environments. AVs feature several sensors and connectivity that can help them make real-time decisions. We propose a novel idea of using autonomous emergency vehicles (AEVs) that can meet the critical response time and drive through a complex road network in smart cities efficiently and safely. This is achieved by considering traffic network analogous to real-time systems (RTS) where we use mixed-criticality real-time system (MCRTS) task scheduling to schedule AEVs for meeting response time.*

## 1. INTRODUCTION

The evolution of present cities into smart cities of the future has provided assurance of easing the way we live. Smart city is mainly focused on urban environment which offers advanced and innovative services to inhabitants to improve the quality of life using information communication technology (ICT) [1]. These advanced and innovative services will help us in solving several current problems easily like traffic congestion, digital security, mobility etc. that are hard to solve using existing technologies. Having impacts on different dimensions, road congestion is one of the major challenges urban planners, traffic authorities and communities are struggling to address. Among different impacts of road congestion, increased response time of emergency vehicles (EVs) like ambulance, fire, police etc. is most severe as it can have an irreparable loss in terms of life and property. According to [2], in the USA only, a delay of one minute in response time increases mortality by 1% which leads to a 7 billion dollars increase in healthcare expenses yearly. To solve the underlying traffic management problem and overcome losses caused by increased congestion, we need advanced ICT-based solutions. Smart cities especially smart transportation provide an ideal environment to implement such solutions.

It is intuitively understood that EVs must get preference over other vehicles when they are traveling to the response scene. EVs get priorities by using special color, sirens and strobe lights, a dedicated green light on approaching traffic signals, special lanes etc. and they travel to service an emergency in an optimized route. To measure the effectiveness of optimization and pre-emption techniques, emergency management services companies are set with a target time to respond to different level of emergencies. For example, St. John's of New Zealand categorizes life-threatening alerts as purple and red, and less threatening events as orange. The contractual target of the purple and red incident is to respond to 50% of cases within 8 minutes and 95% within 20 minutes [3]. In the UK and Canada the target is 75% of purple and red cases within 8 minutes [4], 90% of similar cases within 8 minutes 59 seconds in the USA [5], 50% of cases within 10 minutes in Australia [6], and 92% of cases within 12 minutes in Hong Kong [7].

Over the years, there has been no significant decrease in EV response time [8] because contemporary traffic networks constitute multiple hurdles to the timely movement of EVs. For instance, synchronized operation of traffic lights, increased pedestrian population over cities, continuous construction over lanes and prominently congested road networks have regularly obstructed smooth movement of EVs. In addition, 90% of EVs accidents are caused by human errors. The safety and effectiveness of EVs' movement can be improved if we have access to dynamic road parameters like road congestion, pedestrian flow, travelling time, men at work, halt at road and queued vehicles in real-time and they can be processed to make intelligent driving decisions.

There has been a massive investment in smart cities both from the public and private sectors. large ICT business leaders like IBM, Intel, Siemens, CISCO and SAP are putting a huge effort in developing revolutionary concepts for smart cities. Not only these companies but also governments, philanthropic organizations, and academics are advocating for smart cities. The global smart city market is expected to be valued at US$1.565 trillion in 2020 [9] and the number of smart cities to be 88 by 2025 [10]. As current technology seems insufficient and growth of smart cities is inevitable there is an immediate need to develop ICT driven EV route optimization and pre-emption techniques to meet overwhelming interest and investment.

Smart cities are built on the idea of deep connectivity. Vehicles have access to vehicle-to-X (V2X: vehicle, road, human, infrastructure, internet) communication through several protocols [11]. This connectivity can help in optimizing EV routing. Connectivity gives access to information on dynamic road parameters in real-time. Connectivity also enhances information sharing among smart objects. Present EV routing systems have not blended in real-time traffic data to generate accurate, dynamic and reliable routes for EVs [12]. In the connected environment of a smart city, we can react to dynamic parameters in real-time so that EVs can respond to all levels of emergencies within a certain time. The concept of resource allocation and meeting the timing constraint make EV routing analogous to task scheduling in the real-time system (RTS). In addition, emergencies having several levels of criticality with different service times, makes the EV routing problem very close to scheduling in a mixed-criticality real-time system (MCRTS). MCRTSs have tasks with two or more levels of criticality, for example, non-critical, safety-critical and mission-critical. In MCRTS timing parameters like worst-case execution time (WCET) for processes rely mainly on criticality levels [13].

In this paper, we propose a conceptual model of routing EVs in smart cities. Routing EVs and task scheduling in MCRTS are considered analogous. We use design-by-analogy approach [14] to convert traffic network parameters into MCRTS parameters using different task functions. This allows us to use sophisticated task scheduling algorithms developed for complex MCRTS like in aircraft systems for EV routing. In addition, we have designed this model to route autonomous vehicles (AVs) to serve emergencies. These kind of AVs are termed as autonomous emergency vehicles (AEVs) [15]. So the model is based on a novel idea of routing EVs using task scheduling in MCRTS for autonomous emergency vehicles (AEVs) that can meet the critical response time and drive through a complex road network in smart cities efficiently and safely. Since this is a special case of implementation of emerging technologies like autonomous vehicles and smart cities there is not sufficient literatures to compare our approach with existing approaches.

The approach discussed in the preceding paragraph has multiple contributions for researchers and industry partners. This approach:

1) explores the idea of using AVs in normal mode and emergency mode. The use of AEVs increases traffic safety and connectivity, and these are described in Sec. 2.

2) provides an insight that routing of EVs/AEVs can be done using modern scheduling algorithms developed for MCRTS. For this, it presents an analogical mapping framework in Sec. 3.

3) suggests using dynamic optimization method to find routes for AEVs in smart-cities leveraging access to real-time traffic data in Sec. 4.

4) focuses on multiple levels of emergencies having a different response time. Using MCRTS helps emergency management services to meet or exceed the minimal contractual standards of response time.

5) provides a detailed view of how users, AEVs and real-time traffic management systems communicate with each other.

## 2. AUTONOMOUS EMERGENCY VEHICLES (AEVS)

Autonomous vehicles can sense their surroundings and can move with no or very little human interference [16]. A central computer within the AV analyzes and processes the information received from sensors like global positioning system (GPS), light detection and ranging (LIDAR), video cameras, radar, ultrasonic sensors and then controls steering, brakes, and accelerator in accordance with the formal and informal rules of the road. With the dedicated short range communications (DSRC) system it can communicate with its surroundings.

AVs that are used to serve emergencies are termed as Autonomous emergency vehicles (AEVs). There can be two categories of AEVs. First normal AVs which can also serve for emergencies of lower criticality and second custom-designed AVs e.g. autonomous ambulances. These kinds of AEVs have facilities built within to serve a particular purpose like autonomous ambulances have paramedic facilities. The distinguished property of EVs is that they get priority when they move. AEVs can also get priority by requesting for lane reservation, continuous green light, change of speed limit etc. For this, they are equipped with different communication transmitters and receivers like DSRC, 5G networks etc.

Using of AEVs in place of the traditional emergency service vehicles have the following benefits:

- Use of AEVs will reduce response time and meet or exceed the minimal contractual standards.

- According to National Highway Traffic-Safety Administration only in the USA there is an average of 4500 accidents involving ambulances each year, 3160 accidents involving fire vehicles and 300 fatalities during police pursuit [17]. Use of AEVs improves safety on roads. Fewer crashes as they are without driver error [18].

- Processing of traffic network data allows AEVs to avoid congestion which in turn contribute to lesser carbon emissions due to fuel burning [19].

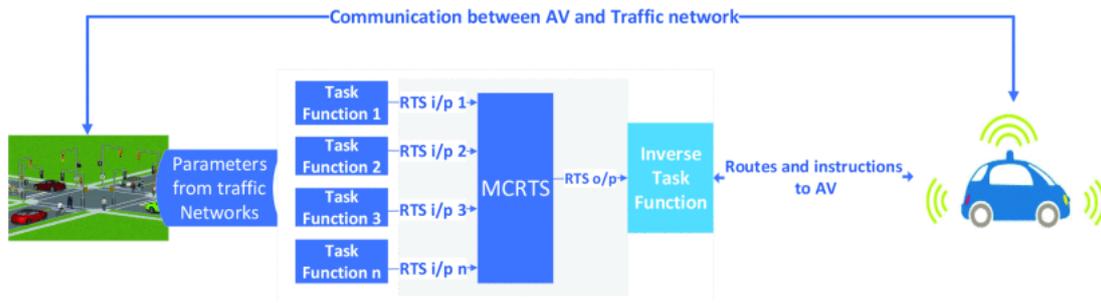

Fig. 1. MCRTS design diagram.

- Provide better mobility to the elderly, youth and children [20].
- AEVs are able to generate and request pre-emption like creating a green wave, lane reservation, informing other vehicles, changing speed limit, use of reverse lane with minimal or no disturbance to other traffic using V2X communication technology [21].
- Locating, instructing and tracking is easier as they are always connected [21].
- Use of AEVs reduces massive expenses in infrastructures like traffic lights, lanes, instructions etc. as these things will be stored in the memory of the vehicle and can be utilized virtually [22].

Due to the higher level of connectivity among the infrastructures, physical infrastructures presently treated as barriers in solving traffic problems can be used like functions which can return values whenever we require. In such condition, the major aim of emergency traffic management system is to align all infrastructures in such a way that emergency vehicles moving within these connected traffic network can respond to emergencies within a predefined time. This means in smart cities with AEVs, the routing of emergency vehicles present complex infrastructure problems that can be converted into timing problems. This provides us with an opportunity of solving routing of emergency vehicles as MCRTS task scheduling problem because the success or failure of MCRTS is completely dependent on solving a task within a stipulated time. The following section explains the relevance of AEV routing in smart cities using MCRTS analogy.

## 3. MIXED CRITICALITY REAL TIME SYSTEM CONCEPT

In emergency response systems used presently, EVs are located at a certain location. Once there is an emergency call, the level of emergency is determined, and response time is set. The number of available EVs with their location is identified. From the present location of the EV to the destination there may exist multiple routes. The optimization system must provide the fastest route to serve within a definite period considering different factors associated with the particular route that may create hindrances in the movement of EV. Next, the system schedules the EV to serve the emergency. If it becomes difficult to respond within the stipulated period due to changing road parameters, the system must be able to provide an alternate route or activate pre-emption to provide priority to EVs so that they can respond to the emergency on or before the set time.

Producing a physical result within a certain time is the basic property of real-time systems (RTS). Inputs from sensors are taken at a periodic interval and real-time computer must send responses to actuators within the chosen time. The ability of the system to produce the results within this chosen time is completely dependent on the system's ability to process necessary computations within time. In case of concurrent events the system must schedule the computations to complete within time. Every task in RTS bears a timing property within which it needs to be processed. While scheduling any task this timing property must be considered by RTS. Therefore, in RTS the accuracy does not only depend on logical results from computation but also on the time when these results are produced. System failure occurs when the system cannot meet this timing property. Therefore, it is indispensable to guarantee the timing property of the system. To guarantee timing behavior, the system must be predictable which means once a task is activated, we must be able to determine its completion time with full confidence [23]. A real-world RTS is usually composed of multiple tasks with multiple criticality levels. If the system fails to meet the timing constraints, we must designate some level of assurance against failure depending upon the criticality of the task. This kind of RTS are termed as mixed-criticality real-time systems (MCRTS) [24].

From the above discussion, we can conclude that there exist certain similarities between routing of EVs and task scheduling in mixed-criticality RTS. There is only one difference between these two approaches. In contemporary emergency response systems, EVs and surrounding cannot communicate with the environment except the use of light strobes and sirens but in smart cities, all the components relate to each other and they can communicate. Smart city provides a connection platform where all the homogeneous smart object communicate using prescribed communication standards. Utilization of interacting traffic resources to process the task of responding to different level of emergencies within precalculated time is like task scheduling in MCRTS. Modeling AEVs routing using MCRTS task scheduling meets the following goals:

- Meeting timing constraints of different emergency responses with different level of criticality.
- Emergency vehicles meet the target response time utilizing available resources but ascertain that other vehicles also make the optimum use of the same resource.

- Though pre-emption is activated it causes the nominal effect to other vehicles.
- Reducing the communication cost between the components of the traffic network system.
- Considers all level of emergencies in terms of critical tasks.
- The scheduling behaviour of EVs in real-time needs to be intelligent, dynamically adaptive, reflexive and reconfigurable.

Parameters of MCRTS are dependent on the level of criticality of each task. Estimates of worst-case execution time (WCET) for any task is also dependent on the level of criticality. For example, the same task can have a lower WCET target if it is assigned as a safety-critical task rather than mission-critical or non-critical tasks [13]. This attribute of MCRTS align completely with our AEVs routing problem where we have different level of emergencies with corresponding response times. In the following subsection we have discussed analogical mapping between MCRTS and traffic network parameters.

**A. MCRTS and Traffic Network Analogy**

Generation of creative ideas for design and problem solving can be sometimes interpreted from the similarity of products, shared functionality or shared relation between items of different domains. This kind of design methodology is termed as design-by-analogy [14]. AEV routing and task scheduling in MCRTS have certain similarity. A MCRTS is a system which provides a certain level of assurance against system failure for some critical tasks. This kind of system exactly matches with the design of the system where AEVs can be used to respond to different level of emergencies. A certain time is allocated for AEVs to respond to a certain emergency case. If it can be responded within that time a task success is noted. In any other case, the system provides some flexibility to the timing constant so that more resources can be assigned to complete the task within WCET. Usually, a MCRTS system comprises of multiple inputs and outputs. Some of them have been listed in Table I.

MCRTS generates outputs like assigning task to processor or assign new deadline considering input parameters like number of available processor or completion time of the present task. There are algorithms to achieve this. With analogical mapping we convert real world traffic network parameters into equivalent MCRTS parameters. This can be achieved by using tasks functions as shown in Figure 1. We use compositional analogy for mapping of traffic network variables with variables of MCRTS. It first does mapping at the level of structure, and that mapping at this level transfers some information. That in turn allows to transfer information at the behavioural level. Once information at behavioural level is transferred, it climbs up this abstraction hierarchy, and transfers information at a functional level [25].

For example, real-world traffic network have inputs like EV location, destination, possible routes, congestion level of road network, previously selected route, halt on road, speed limit of the road, number of lanes, likely speed, travel time from previous user, time of day, slope on road, number of traffic nodes, roundabout, traffic lights, pedestrian flow, queued vehicles, recovery time of traffic pre-emption, traffic phase timing etc. Task functions use analogical mapping to maps these traffic domain input parameters to MCRTS inputs. Now this allows us to use properties of MCRTS. The outputs of MCRTS is now passed through inverse task functions which finally convert MCRTS output into equivalent traffic parameters. For example, assign processor can be equivalent to assign route.

Table I. INPUTS AND OUTPUTS IN MCRTS

| Inputs | Outputs |
|---|---|
| Number of periodic, aperiodic or sporadic tasks | Assigning task to processor |
| Number of Pre-emptive and non-pre-emptive tasks | Assign new deadline |
| Number of Fixed or dynamic priority tasks | Queue task |
| Number of Independent or dependent tasks | Alter priority |
| Number of processors | Assign pre-emption etc. |
| Number of reserved processor | |
| Release time, completion time deadlines, priority, precedence, constraints | |

In the following section, we have elaborated the concept of modelling AEVs routing using MCRTS task scheduling discussed in section 1, section 2 and section 3 using mathematical notations and different diagrams.

## 4. A CONCEPTUAL MODEL OF THE SYSTEM

In this section, we are introducing a novel model of route optimization and pre-emption for different types of AVs. We assume that the service function $S(e, c, r, v, p)$ is the set of all the attributes required to route an AV in AEV mode from a source to destination. A simplified diagram to visualize the process of how AVs service is represented in Figure 2. Here, $e$ stands for level of emergency, $c$ stands for the level of criticality, $r$ represents the number of available routes, $v$ represents the number of available autonomous vehicles and $p$ stands for the type of pre-emption to be activated. The process is defined below:

- Mode ($E_i$) represents the service mode of AVs. It can take two values $E_o$ and $E_1$. $E_o$ represents AVs are serving in normal mode and $E_1$ represents AVs are operating on AEV mode. These values are updated by the user who requests the AV service.

- Criticality ($C_i$) is level of criticality of emergency that AVs are going to serve. From normal practice in different countries, we can have a total of four values of criticality $C_o \ldots, C_3$. $C_o$ represents no critical emergency case so AVs can operate in normal mode. $C_3$ and $C_2$ are life-threatening alerts that are symbolised as purple and red by emergency management service companies and $C_1$ are orange cases which are less life-threatening. These values are updated by the user by answering certain questions that appear in their application.

- Route ($R_i$) is the number of available routes from source to the destination of the service. It can take any values from $R_o \ldots \ldots, R_n$ depending upon the number of traffic nodes available in that particular geographical location. The optimization function calculates the fastest path considering all the road parameters and supply the value to the system.

- Vehicle ($V_i$) represents a number of available AVs to serve. $V_o \ldots, V_n$ are the possible types of AVs. AVs are normal autonomous car that can also serve in cases of less critical emergencies like user needs to visit hospital and doesn't require any paramedic support during travel. Other AVs can be autonomous ambulance, fire, police car etc. GPS system installed inside AVs and their service notification status give the value of ($V_i$).

- Pre-emption ($P_i$) represents the instruction to the AVs weather to activate pre-emption or not. It also instructs which type of pre-emption to activate like creating a green wave, lane reservation, informing other vehicles, changing the speed limit, use of reverse lane with minimal or no disturbance to other traffic etc. It can take values from $P_o \ldots, P_n$. $P_o$ symbolizes no use of pre-emption and all other values are the type of pre-emption the system suggests to activate.

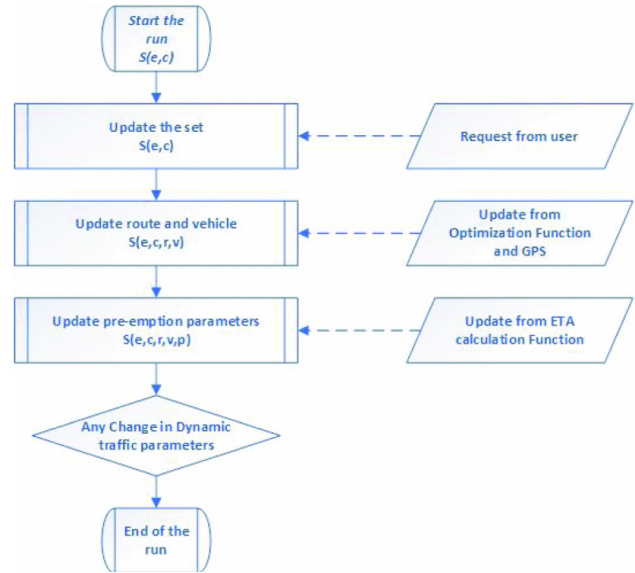

Fig. 2. Execution of service for AEVs

The suggestion of pre-emption is guided from the calculation of estimated arrival time (ETA). As we are suggesting MCRTS approach, we can compare this ETA with WCET for AVs to serve. Once all the values of service function $S(e, c, r, v, p)$ is calculated the user and AVs are advised with the service time. During the service, if there is any change in the dynamic traffic parameters like pedestrian flow, congestion, traveling time, men at work, halt at road, queued vehicles etc. a new updated ETA are advised to user and AVs with proper pre-emption instruction.

The core of our conceptual model is the visualization of AVs as AEVs. User using a simple mobile app can initiate this service and most of AVs can serve in different types of emergencies as AEVs. The system is dynamic and keeps on updating in real time with the help of data received from the array of sensors installed inside AVs and environment. The entire system has five components mobile application for users, AV sensory system, AV control system, dedicated short-range communication (DSRC) system and real-time traffic control system. All these systems and their interactions are shown in Figure 3.

The solution provided has five major components:

• *Mobile Application for user:* This component is focused on user side application where different users request AV's service. They can request for any kind of emergency service. Their request sets AVs to operate in AEV mode. Their responses to certain questions can also set the criticality level of the emergency, source, and destination of the service and also the type of EV to be scheduled to serve the emergency.

• *AV sensory system:* This comprises of different sensors that AV is installed with like LIDAR, radar, ultrasonic sensors, GPS, video camera etc. These all sensor help AV to visualize the environment.

• *AV control system:* AV's central computer processes all the sensor data and instructions received from real-time traffic management system, processes it and generates driving instructions for motor, steering, and braking.

• *Dedicated Short-Range Communication (DSRC) system:* This system permits AV to establish V2X communication using different transmitters and receivers.

• *Real-time traffic control system:* The core idea of our conceptual framework lies within real-time traffic control system. The system gets service mode information ($E_i$) and level of criticality ($C_i$) and type of vehicle ($V_i$) to deploy to serve from the user using a mobile application. Once these parameters are known the system initiates the optimization function and returns the route with its associated ETA. This ETA is now set as WCET of the MCRTS. All the available resources are now assigned to execute the task of responding to the service (emergency or non-emergency) within WCET.

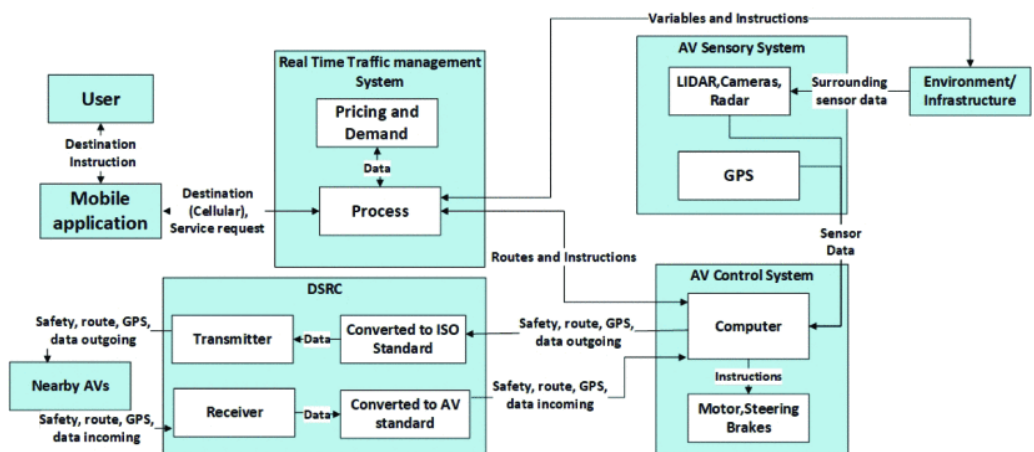

Fig. 3. System interface diagram

## 5. CONCLUSION

In this paper we introduce a conceptual model of routing autonomous emergency vehicles to respond to emergencies using a mixed-criticality real-time systems (MCRTS) approach in smart cities. To use the highly refined scheduling algorithms of complex MCRTS we have suggested the use of analogical mapping between traffic network parameters and MCRTS. The preliminary goal of this paper is to use autonomous vehicles as part of an emergency response system in smart cities termed as autonomous emergency vehicles. Through an analogical mapping between MCRTS and the AEV routing problem, we propose a framework to route AEVs using dynamic traffic parameters.